\documentclass[aps,pre,groupedaddress,twocolumn,showpacs]{revtex4}

\usepackage{graphics}
\usepackage{amsmath}
\usepackage{amsfonts}
\usepackage{amssymb}
\usepackage{fancyhdr}

\renewcommand{\vec}[1]{\mbox{\boldmath$#1$}}
\newcommand{\pr}{\|}

\newcommand{\0}{0}

\newcommand{\lmode}{\ell}

\begin{document}

\title{Fluctuation spectrum of quasispherical membranes with force-dipole activity}

\author{Michael A. Lomholt}
\email[]{mlomholt@memphys.sdu.dk}
\affiliation{NORDITA - Nordic Institute for Theoretical Physics, Blegdamsvej 17, 2100 Copenhagen \O, Denmark}
\affiliation{The MEMPHYS Center for Biomembrane Physics, University of Southern Denmark, Campusvej 55, DK-5230 Odense M, Denmark}

\date{\today}

\begin{abstract}
The fluctuation spectrum of a quasi-spherical vesicle with active membrane proteins is calculated. The activity of the proteins is modeled as the proteins pushing on their surroundings giving rise to non-local force distributions. Both the contributions from the thermal fluctuations of the active protein densities and the temporal noise in the individual active force distributions of the proteins are taken into account. The noise in the individual force distributions is found to become significant at short wavelengths.
\end{abstract}


\maketitle


\section{Introduction}
Biological membranes are mixtures of basically lipids and proteins that actively participate in many biological processes. The lipids, being the basic component, form a bilayer in which proteins are included. Many of these membrane proteins are molecular machines, which can use some readily available energy source, for instance light or ATP, to perform different tasks \cite{alberts02}. A prominent example is the transportation of ions across a membrane against a chemical potential gradient. Due to this constant activity a biological membrane is a system out of thermal equilibrium.

One way to observe non-equilibrium behavior of membranes is to study their shape fluctuations. Lipid membranes in their fluid phase are flexible structures undergoing thermally excited shape fluctuations that can be observed directly by video microscopy \cite{seifert97}. Indirect studies of the fluctuations, using micromanipulation techniques to measure the amount of membrane excess area stored in the fluctuations, have already given evidence that activated proteins incorporated in lipid membranes enhance these fluctuations \cite{manneville99,girard05}. Efforts are being put into extending these studies to use video microscopy to obtain the fluctuation spectrum directly \cite{pecreaux04}.

A theoretical model was proposed in \cite{manneville01}, which explains the activity induced enhancement of the fluctuations as being a result of the active proteins pushing on their fluid surroundings. This pushing could arise when a protein is pumping material across the membrane, or when it is changing its configuration as part of an active cycle. This model will be called the force-dipole model here, because it was formulated mathematically as each protein contributing with a force-dipole to the hydrodynamic equations of the fluids surrounding the membrane. The consequences of this model for excess area stored in the fluctuations of a planar membrane was calculated in \cite{manneville01}, and from this calculation it was found that the model constitutes a possible explanation of the results of the micromanipulation experiments.

A video microscopy study of quasi-spherical vesicles would test the force-dipole model further, because the fluctuations at different wavelength can be measured, and not just the sum over all wavelength giving the excess area. However, it is not easy to calculate the predictions of the model for the fluctuation spectrum of a quasi-spherical vesicle directly, because the equations of motion for the surrounding bulk fluids would then have to be solved in a spherical geometry including the presence of the force-dipoles.

In \cite{lomholt05c} the force-dipole model was reformulated into an equivalent but calculationwise more manageable formulation, which was called the Gibbs formulation for the system. One purpose of the present paper is to show how this reformulation makes it possible to calculate the fluctuation spectrum of a quasi-spherical vesicle.

A second purpose of this paper is to include the non-thermal noise of the activity in the calculation of the fluctuation spectrum. This was neglected in \cite{manneville01}, where the magnitude of the force-dipole generated by each active protein was treated as being constant in time. However, there does not seem to be good reasons to assume that the forces with which a protein is actively pushing on its surroundings is constant in time as the protein moves through one cycle of, say, pumping an ion across the membrane. Estimates performed for a planar membrane in \cite{gov04}, where the effect of active noise were treated independently of the activities time-averaged constant effect on the dynamics, showed that the noise can have a significant effect. Here we will include both effects for force-dipoles, the time averaged and the corresponding temporal noise, and calculate the combined effect on the fluctuation spectrum of a quasi-spherical vesicle.

The presentation of the calculation will begin in section \ref{sec:neareqdyn}, where the near-equilibrium dynamics of the membrane is established. The dynamics is formulated as a set of Langevin equations in section \ref{sec:thermnoise}, and the appropriate statistics of the thermal noise is worked out. This is in turn based on a calculation of the equilibrium fluctuations performed in appendix \ref{sec:appfluc}. The activity is introduced in section \ref{sec:activity}, where the Langevin equations for the membrane dynamics is modified to include the active force-dipoles. The main result of the paper, equation (\ref{eq:anaform}), is in section \ref{sec:flucspec}, where the activity modified Langevin equations are solved for the shape fluctuations of the vesicle and the different contributions to the result are estimated and discussed. In section \ref{sec:microexp} it is briefly sketched what impact the inclusion of the non-thermal noise has on the analysis of the existing micropipette experiments. Finally, a conclusion is given in section \ref{sec:concl}.

\section{Near-equilibrium dynamics}\label{sec:neareqdyn}
In this section we will set up the near-equilibrium dynamics following the formalism developed in \cite{lomholt05}. The notational conventions regarding differential geometry is the same as in \cite{lomholt05,lomholt05c}.

Our starting point is an excess Helmholtz free energy $F$ of the membrane which is
\begin{align}
F=\int_{\rm M} d A\;\big[&2\kappa H^2+\Lambda\left(n_{p^+}-n_{p^-}\right)H\nonumber\\
&-\sigma_\phi(\phi-1)+\phi V(n_{p^+}/\phi,n_{p^-}/\phi)\big]\;.\label{eq:free}
\end{align}
The first term re\-pre\-sents the bending resistance of the membrane with $\kappa$ being the bending rigidity and $H$ the mean curvature. The second term is a coupling between the mean curvature and the protein concentrations per membrane area, $n_{p^+}$ and $n_{p^-}$, where $+$ and $-$ label the two possible orientations that an asymmetric transmembrane protein in a membrane can have. The third term is a consequence of the observation that inhomogeneities arising from local compression of the area of the membrane relax much faster than bending deformations \cite{miao02}. The term therefore has the form of a local constraint with $\sigma_\phi$ being a Lagrange multiplier field \footnote{The reader might wonder if we are not overly complicating the analysis here in comparison with \cite{manneville01}. The reason that we have to consider lateral compression of the membrane here is that for spherical geometries the bulk hydrodynamics will couple the lateral two-dimensional flow of the membrane fluid to the membrane motion in the membranes normal direction. This is not the case for planar membranes.}. The field $\phi$, which is constrained to be $1$, is a linear combination of the different density fields $n_A$ of molecules in the membrane, with each species (including both proteins and lipids) weighted by the area $a^A$ the individual molecules prefer to take up in the membrane
\begin{equation}
\phi=\frac{1}{1+\alpha}\sum_A a^A n_A\;.\label{eq:phidefini}
\end{equation}
The unitless number $\alpha$ is a global quantity that measures the mismatch between the area that the membrane prefers to have,
\begin{equation}
A_{\rm preferred}=\sum_A a^A \int_M d A\; n_A\;,
\end{equation}
and the actual area of the membrane
\begin{equation}
A=\int_{\rm M} d A\;.
\end{equation}
It is explicitly
\begin{equation}
\alpha=\frac{A_{\rm preferred}-A}{A}\;.\label{eq:Apconstr}
\end{equation}
The average of $\sigma_\phi$ over the area of the membrane is thermodynamically conjugate to the area $A$, and thus this average can be understood as the tension of the membrane. The fourth term in Equation (\ref{eq:free}) is an effective interaction potential between the protein fields $n_{p^\pm}$. The functional form is left unspecified, but it should include at least the mixing entropy for the protein fields $n_{p^\pm}$, and could also contain terms related to other interactions between the protein molecules. The $\phi$-field in this term is introduced for convenience, because then the potential $V$ will not be present in the force balance equation of the membrane, see Equations (\ref{eq:frs}) to (\ref{eq:frsa}) later.

Having specified the free energy in Equation (\ref{eq:free}), we can now follow the formalism of \cite{lomholt05} and write down the dynamics of the membrane. At low Reynolds number the equation of motion for the shape field $\vec{R}$ is simply the condition of force balance of the membrane. Following \cite{manneville01} we ignore internal dissipation in the membrane and write the force balance as
\begin{equation}\label{eq:eqforcebal}
\vec{f}_{\rm rs}+\vec{T}^++\vec{T}^-=0\;.
\end{equation}
The force $\vec{f}_{\rm rs}$ is the elastic restoring force per area of the membrane
\begin{align}
\vec{f}_{\rm rs}=&-\frac{1}{\sqrt{g}}\left.\frac{\delta F}{\delta \vec{R}}\right|_{\sqrt{g}n_A,\sigma_{\phi}} \equiv f_{\rm rs,n}\vec{n}+f_{{\rm rs},\alpha}\vec{t}^\alpha\;,\label{eq:frs}\\
f_{{\rm rs,n}}=&2H\sigma_\phi-4\kappa H\left(H^2-K\right)-2\kappa\vec{\nabla}_\pr^2H\nonumber\\
&-\Lambda \left(2 H^2-K\right)n_\Delta-\frac{\Lambda}{2}\vec{\nabla}_\pr^2n_\Delta\;,\\
f_{{\rm rs},\alpha}=&\partial_\alpha\sigma_\phi-\Lambda n_\Delta\partial_\alpha H\;,\label{eq:frsa}
\end{align}
where we have introduced the density difference field $n_\Delta=n_{p^+}-n_{p^-}$. This field has a natural partner in the field $n_\Sigma=n_{p^+}+n_{p^-}$, which will be used later. The force densities $\vec{T}^\pm$ are the stresses of the bulk fluids on the membrane surface. At low Reynolds number the bulk fluids are governed by the Navier-Stokes equation
\begin{equation}
\eta \vec{\nabla}^{2} \vec{v}_\pm-\vec{\nabla} p^\pm=0 \;,\label{eq:noactNS}
\end{equation}
together with the incompressibility condition $\vec{\nabla}\cdot\vec{v}_\pm=0$. Here $\vec{v}_\pm=\vec{v}_\pm(\vec{r},t)$ are the velocities of the bulk fluids with $\vec{r}$ labeling the position and $t$ time. $p^\pm$ are the pressure fields and $\eta$ the viscosity. Given a solution of the bulk hydrodynamics we can calculate the stress tensors of the bulk fluids as
\begin{equation}
\mathsf{T}^\pm=-p^\pm\mathsf{I}+\eta\left[\vec{\nabla}\vec{v}_\pm+\left(\vec{\nabla}\vec{v}_\pm\right)^T\right]\;,
\end{equation}
where $\mathsf{I}$ is the identity tensor. Then the forces $\vec{T}^\pm$ are
\begin{equation}\label{eq:Tvecdef}
\vec{T}^\pm=\pm \vec{n}\cdot\left.\mathsf{T}^\pm\right|_{\vec{r}=\vec{R}}\;.
\end{equation}
To actually find a solution of the bulk hydrodynamics we need boundary conditions for the bulk velocity fields $\vec{v}_\pm$ at the membrane surface. The boundary condition that is usually imposed when dealing with viscous fluids is the so-called no-slip boundary condition, where the bulk fluid at the boundary is constrained to have the velocity of the boundary material. In \cite{lomholt05} it was argued that for a multicomponent membrane the general form of this boundary condition was a linear combination of the flows of the different species of molecules in the membrane. If we let $j^\alpha_A$ be the current for the density $n_A$ entering the conservation law \cite{lomholt05}
\begin{equation}\label{eq:nAcons}
D_t n_A + D_\alpha j^\alpha_A=0\;,
\end{equation}
then we can write such a linear combination as
\begin{equation}\label{eq:vbound1}
\left.\vec{v}_\pm\right|_{\vec{r}=\vec{R}}= \partial_t\vec{R}+\sum_A L^A_\pm j^\alpha_A\vec{t}_\alpha\;.
\end{equation}
where the phenomenological constants $L^A_\pm$ are areas of molecular sizes. They should satisfy the constraint $\sum_A L^A_\pm n_A=1$ such that it is possible to have uniform motion of the whole system. In \cite{lomholt05} it was argued that the precise values of the $L^A_\pm$ can influence the distribution of proteins in the membrane if the membrane is subjected to a bulk shear flow with velocities of the order $k_B T/(\eta L^A_\pm)$. However, the velocities in the bulk fluids induced by thermal noise are of the order $k_B T/(\eta \lambda^2)$, where $\lambda$ is the wavelength of the undulation in question. Thus when we are interested in undulations much longer than the molecular length scales we can ignore shear flow effects arising due to thermal fluctuations. This means that the detailed values of the $L^A_\pm$ are unimportant for the coupling of the bulk fluid dynamics to the dynamics of the protein densities in the present problem. Thus we are allowed to make a convenient choice that will simplify calculations. We will choose: $L^A_\pm=a^A/(1+\alpha)$ such that
\begin{equation}\label{eq:vbound}
\left.\vec{v}_\pm\right|_{\vec{r}=\vec{R}}= \partial_t\vec{R}+\frac{1}{1+\alpha}\sum_A a^A j^\alpha_A\vec{t}_\alpha\equiv \vec{v}_\phi\;.
\end{equation}
A reason that the boundary condition in Equation (\ref{eq:vbound}) is convenient can be found by multiplying (\ref{eq:nAcons}) by $a^A$, summing over $A$ and using the constraint $\phi=1$ to find
\begin{equation}
\frac{\partial_t\sqrt{g}}{\sqrt{g}}=-\frac{\dot \alpha}{1+\alpha}-D_\alpha v^\alpha_\phi\;,\label{eq:phicons2}
\end{equation}
where $v^\alpha_\phi\vec{t}_\alpha=\vec{v}_\phi-\partial_t\vec{R}$. Thus the motion of the membrane shape $\vec{R}$ is related to the motion of the bulk fluids $\vec{v}_\pm$ at the membrane surface, without the protein densities $n_{p^\pm}$ entering directly.

From Equations (\ref{eq:frs}) to (\ref{eq:frsa}) we see that the dynamics of the shape field $\vec{R}$ is coupled to the density difference field $n_\Delta$. The equation of motion for $n_\Delta$ is a conservation law, Equation (\ref{eq:nAcons}), but we need the expression for the current $j^\alpha_\Delta$. This current can be divided into two contributions, $j^\alpha_\Delta=n_\Delta v^\alpha+j^\alpha_{\Delta,{\rm d}}$, where the first term is the convective flow with $v^\alpha=\sum_A m^A j^\alpha_A/(\sum_B m^B n_B)$ representing the kinematic velocity of the flow in the membrane and $m^A$ being the mass of a molecule of species $A$. The second term is the dissipative part of the current. Close to equilibrium we can write down a constitutive relation coupling this current linearly to possible driving forces. The only driving force we have that transform in the same way as $j^\alpha_A$ under symmetry transformations (including the approximate symmetry of exchanging $n_{p^+}$ and $n_{p^-}$) is the gradient of the corresponding chemical potential \cite{lomholt05}
\begin{equation}
\mu^\Delta=\left.\frac{\delta F}{\delta n_\Delta}\right|_{\vec{R},n_\Sigma,\phi=1}=\left.\frac{\partial V}{\partial n_\Delta}\right|_{n_\Sigma,\phi=1}+\Lambda H\;.
\end{equation}
Therefore we can write the constitutive relation as
\begin{equation}
j_{\Delta,{\rm d}}^\alpha=-\Omega_{\Delta\Delta}\partial^\alpha\mu^\Delta\;,
\end{equation}
where $\Omega_{\Delta\Delta}$ is a phenomenological parameter related to the diffusion constant of the proteins, see Equation (\ref{eq:Deltadiff}) later.

Having worked so far with equations that are true for any membrane shape, we will now restrict ourselves to a shape which is fluctuating around a spherical average shape with radius $R_0$, and only keep terms up to first order in deviations from this average shape. We expand in spherical harmonics $\mathcal{Y}_{\ell m}=\mathcal{Y}_{\ell m}(\theta,\phi)$, which are eigenfunctions of the Laplace operator on the sphere with eigenvalues $-\ell(\ell+1)/R_0^2$, and they are orthonormal
\begin{equation}
\int_0^{2\pi}d \phi\int_0^\pi\sin\theta\,d\theta\;\mathcal{Y}_{\ell' m'}^*\mathcal{Y}_{\ell m}=\delta_{\ell'\ell}\delta_{m'm}\;.
\end{equation}
The expansion will be written as
\begin{align}
\vec{R}&=\vec{R}_0+\vec{n}_0 R_0\sum_{\lmode,m} u_{\ell m}\mathcal{Y}_{\ell m}\;,\\
n_\Delta&=\sum_{\lmode,m} n_{\Delta,\ell m}\mathcal{Y}_{\ell m}\;,\\
n_\Sigma&=n_{0,\Sigma}+\sum_{\lmode,m} n_{\Sigma,\ell m}\mathcal{Y}_{\ell m}\;,\\
\sigma_\phi&=\sigma_{0,\phi}+\sum_{\lmode,m} \sigma_{\phi,\ell m}\mathcal{Y}_{\ell m}\;,
\end{align}
where the subscript $0$ indicates that the quantity has the value of the average state and $u_{\ell m}$, $n_{\Delta,\ell m}$, $n_{\Sigma,\ell m}$ and $\sigma_{\phi,\ell m}$ represent small deviations from this state. Note that we follow \cite{manneville01} in making the simplifying assumption that the proteins are distributed symmetrically between the two possible orientations in the membrane such that $n_{0,\Delta}$ vanishes.

The solution of the equations of motion to zeroth order in the deviations from a spherical membrane shape is that $R_0$ and $n_{0,\Sigma}$ are constant due to volume and molecular number conservation. Furthermore the force balance condition, Equation (\ref{eq:eqforcebal}), becomes
\begin{equation}
2H_0\sigma_{0,\phi}+p_0^--p_0^+=0\;,\label{eq:laplacelaw}
\end{equation}
which is Laplace's law with $p_0^\pm$ being the average pressures in the bulk. The value of the tension $\sigma_{0,\phi}$ is in principle related to the value of $\alpha$, but we will choose $\sigma_{0,\phi}$ to be the free parameter here. The reader is referred to \cite{henriksen04} for further details on the subject of membrane tensions and their conjugate areas.

Calculating to first order in the deviations from a spherical shape is more tedious. Especially we need to find expressions for the hydrodynamic forces $\vec{T}^\pm$ in terms of the state variables of the membrane. Fortunately, a solution to the Navier-Stokes equation at low Reynolds number in spherical geometry is known. This solution is called the Lamb solution \cite{kim91}, and we can use it to find $\vec{T}^\pm$ in terms of $\vec{v}_\phi$. Having worked this out we can then eliminate $\sigma_{\phi,\ell m}$ and $v^\alpha_\phi$ from the dynamics of $u_{\ell m}$ and $n_{\Delta, \ell m}$ to arrive at a closed set of first order differential equations for $u_{\ell m}$ and $n_{\Delta, \ell m}$. $n_{\Sigma,\ell m}$ will not enter the equations for $u_{\ell m}$ and $n_{\Delta, \ell m}$, because $\partial^2 V/(\partial n_\Delta \partial n_\Sigma)|_{n_\Delta=0}=0$ since the system should then be symmetric with respect to reversing the two sides of the membrane. We do not write out the intermediate results of the full calculation here. Instead we refer to \cite{miao02} where an analogous calculation was performed in detail. We state the final result by collecting the dynamical fields into a column vector $\vec{w}_{\ell m}$, where the transpose of $\vec{w}_{\ell m}$ is $\vec{w}_{\ell m}^T=\left(u_{\lmode m},n_{\Delta,\ell m}\right)$, and then write the differential equations as
\begin{equation}
B{\vec{\dot {w}}}_{\ell m}=-A\vec{w}_{\ell m}\;.\label{eq:nonnoise}
\end{equation}
Here the matrix $B$ represents the dissipative forces
\begin{equation}
B=\left(\begin{array}{cc}\eta R_0^3\frac{4\ell^3+6\ell^2-1}{\ell(\ell+1)}&0\\0&\frac{R_0^4}{\Omega_{\Delta\Delta}\ell(\ell+1)}\end{array}\right)\;,
\end{equation}
and the matrix $A$ the elastic ones
\begin{equation}
A=\left(\begin{array}{cc}E_\ell&-\frac{\Lambda R_0}{2}(\ell+2)(\ell-1)\\
-\frac{\Lambda R_0}{2}(\ell+2)(\ell-1)& \chi R_0^2\end{array}\right)\;,
\end{equation}
where
\begin{align}
E_\ell&=\left(\kappa\ell(\ell+1)+\sigma_{0,\phi}R_0^2\right)(\ell+2)(\ell-1)\;,\label{eq:El}\\
\chi&\equiv\left.\frac{\partial^2 V}{\partial n_\Delta^2}\right|_{n_{0,\Sigma},n_{\Delta}=0}\;.\label{eq:VDelta}
\end{align}

\section{Thermal noise}\label{sec:thermnoise}
Equation (\ref{eq:nonnoise}) represents the near-equilibrium dynamics of the average shape of the membrane. To include the effect of the thermal forces on the dynamics of the shape we follow the Langevin approach \cite{kampen97} and add a white-noise term $\vec{\eta}$ to the dynamic equations. The result is the stochastic equation
\begin{equation}
{\vec{\dot {w}}}_{\ell m}=-B^{-1}A\vec{w}_{\ell m}+\vec{\eta}_{\rm thermal}\;,\label{eq:noise}
\end{equation}
where the statistics of the noise obeys
\begin{equation}
\left<\vec{\eta}_{\rm thermal}(t)\vec{\eta}^\dagger_{\rm thermal}(t')\right>=\Gamma_{\rm thermal}\delta\left(t-t'\right)
\;,
\end{equation}
with $\vec{\eta}_{\rm thermal}^\dagger$ representing the complex conjugate of $\vec{\eta}_{\rm thermal}^T$. The correlation matrix $\Gamma_{\rm thermal}$ can be determined from the condition that the long time limit of the fluctuations calculated from Equation (\ref{eq:noise}) should equal the fluctuations that can be calculated directly from the free energy using equilibrium statistical mechanics. The equilibrium statistical mechanical calculation is performed in appendix \ref{sec:appfluc} with the result
\begin{equation}
\left<\vec{w}_{\ell m}\vec{w}_{\ell m}^\dagger\right>_{\rm eq.}=k_{\rm B}T A^{-1}\;.\label{eq:eqfluc}
\end{equation}
To find the long time limit of the fluctuations from Equation (\ref{eq:noise}) we first note that Equation (\ref{eq:noise}) has the formal solution
\begin{align}
\vec{w}_{\ell m}(t)=&e^{-(B^{-1}A)t}\vec{w}_{\ell m}(t=0)\nonumber\\
&+e^{-(B^{-1}A)t}\int_0^t dt'\;e^{(B^{-1}A)t'}\vec{\eta}_{\rm thermal}(t')\;.\label{eq:formalsol}
\end{align}
Squaring this and taking the average we find that the value at long times of the fluctuations is
\begin{equation}
\left<\vec{w}_{\ell m}\vec{w}_{\ell m}^\dagger\right>=\int_{0}^\infty dt\;e^{-(B^{-1}A)t}{\Gamma}_{\rm thermal}e^{{-(B^{-1}A)}^T t}\;.\label{eq:steadyfluc}
\end{equation}
Using $B^{-1}(B^{-1}A)^T=(B^{-1}A)B^{-1}$ it can easily be checked that the choice of $\Gamma_{\rm thermal}$ that ensures the agreement between the two ways of calculating the fluctuations is
\begin{equation}
\Gamma_{\rm thermal}=2k_{\rm B}T B^{-1}\;.
\end{equation}

\section{Activity}\label{sec:activity}
Having set up the stochastic equations for the near-equilibrium dynamics we will now modify these to include the effect of the active force-dipoles. Most of the work needed to do this was done in \cite{lomholt05c}, where it was found that the important contribution of the force-dipoles to the dynamics was to modify the force balance condition for the membrane to
\begin{equation}\label{eq:actforce}
\vec{f}_{\rm rs}+\vec{f}_{\rm act}+\vec{T}^++\vec{T}^-=0\;,
\end{equation}
where $\vec{f}_{\rm act}$ is the force per area generated by the activity. In \cite{lomholt05c} this force was obtained as a divergence of a stress tensor
\begin{equation}
\vec{f}_{\rm act}=D_\alpha\left(T^{\alpha\beta}_{\rm act}\vec{t}_\beta+T^\alpha_{\rm n,act}\vec{n}\right)\;,
\end{equation}
where
\begin{align}
T_{\rm act}^{\alpha\beta}&=\sigma_{\rm dip}g^{\alpha\beta}+\left(Hg^{\alpha\beta}+\frac{1}{2}K^{\alpha\beta}\right)Q\;,\label{eq:tangTact}\\
T_{\rm n,act}^\alpha&=\frac{1}{2}\partial^\alpha Q\;.\label{eq:normTact}
\end{align}
The quantity $\sigma_{\rm dip}$ is a modification of the tension induced by the activity. Due to the fast balancing of forces in the tangential directions of the membrane, fluctuations in $\sigma_{\rm dip}$ will immediately be counterbalanced by changes in $\sigma_\phi$, and therefore only the average value $\sigma_{0,{\rm dip}}$ will have an influence on the dynamics of the membrane shape. A discussion of the magnitude of this contribution can be found in \cite{lomholt05c}. $Q$ is a measure of how much the bending moments in the membrane is modified by the activity of the force-dipoles. The model we will use for $Q$ is
\begin{align}
Q&= F_{(2)} n_{\Delta}+S\;.
\end{align}
The constant $F_{(2)}$ is related to the parameters of \cite{manneville01} by being the quadrupole moment of the shifted force-dipole of an active protein:
\begin{equation}
F_{(2)}=\left(\left(w^\uparrow\right)^2-\left(w^\downarrow\right)^2\right)F_a\;.
\end{equation}
Here $F_a$ is a force and $w^\uparrow$ and $w^\downarrow$ are length scales related to how the active proteins insert themselves in the membrane. $S$ represents the noise arising when the force-dipole generated by an individual protein is not constant during the cycle of the activity. It has zero average
\begin{align}
\left<S\right>&=0\;,
\end{align}
and we model its autocorrelation function similarly to \cite{gov04,lin06,prost96,prost98} by combining the self-diffusion of the proteins with an exponential decay of the correlations in the proteins internal state
\begin{multline}
\left<S(\theta,\phi,t)S({\theta'},{\phi'},t')\right>=\\ \Gamma_{\rm a}n_{0,\Sigma}G_{\rm sp}({\theta},{\phi},{\theta'},{\phi'},|t-t'|)e^{-|t-t'|/\tau_{\rm a}}\;.
\end{multline}
where $\Gamma_a$ is a phenomenological constant, $\tau_a$ a correlation time and $G_{\rm sp}$ is the single particle density correlation function for the active proteins. $G_{\rm sp}$ is the solution of the diffusion equation on a spherical vesicle
\begin{equation}
\frac{\partial}{\partial t}G_{\rm sp}({\theta},{\phi},{\theta'},{\phi'},t)=D_{\rm sp}D_\alpha \partial^\alpha G_{\rm sp}({\theta},{\phi},{\theta'},{\phi'},t)\;,
\end{equation}
with the initial condition
\begin{equation}
\left.G_{\rm sp}({\theta},{\phi},{\theta'},{\phi'},t)\right|_{t=0}=\frac{1}{\sqrt{g_0}}\delta({\theta}-{\theta'})\delta({\phi}-{\phi'})\;,
\end{equation}
where $D_{\rm sp}$ is the single particle diffusion constant and the spatial derivatives act on $(\theta,\phi)$.

Note that, contrary to \cite{manneville01,lomholt05c}, we have for simplicity ignored any dependence of the protein activity on the local curvature of the membrane, but included internal noise in the model instead.

We calculate the stochastic equations of motion for $u_{\ell m}$ and $n_{\Delta,\ell m}$ as before using the force balance condition of Equation (\ref{eq:actforce}), instead of Equation (\ref{eq:eqforcebal}). This gives a changed set of stochastic equations which becomes
\begin{equation}
{\vec{\dot {w}}}_{\ell m}=-C\vec{w}_{\ell m}+\vec{\eta}_{\rm thermal}+\vec{\eta}_a\;,\label{eq:noise2}
\end{equation}
with a modified dynamical matrix
\begin{equation}
C=B^{-1}(A+A_a)\;,
\end{equation}
where
\begin{align}
A_a&=\left(\begin{array}{cc}\sigma_{0,{\rm dip}}R_0^2(\ell+2)(\ell-1)&\frac{{F_{(2)}}{R_0}}{2}(\ell+2)(\ell-1)\\0&0\end{array}\right)\;,\label{eq:B1}
\end{align}
and an additional noise term
\begin{equation}
\vec{\eta}_a=B^{-1}\left(\begin{array}{c}R_0(\ell+2)(\ell-1)S_{\ell m}/2\\ 0\end{array}\right)\;,
\end{equation}
where $\left<S_{\ell m}\right>=0$ and
\begin{equation}
\left<S_{\ell m}^*(t)S_{\ell' m'}(t')\right>=\frac{n_{0,\Sigma}\Gamma_a}{R_0^2}e^{-|t-t'|/\tau_\ell}\delta_{\ell\ell'}\delta_{m m'}\;.
\end{equation}
The time scale $\tau_\ell$ is a combination of $\tau_a$ and the relaxation time for the single protein diffusion
\begin{equation}
\frac{1}{\tau_\ell}=\frac{1}{\tau_a}+\frac{D_{\rm sp}\ell(\ell+1)}{R_0^2}\;.
\end{equation}

\section{Fluctuation spectrum}\label{sec:flucspec}
To find the fluctuation spectrum from Equation (\ref{eq:noise2}) we first note that formula (\ref{eq:steadyfluc}) is now modified to
\begin{equation}
\left<\vec{w}_{\ell m}\vec{w}_{\ell m}^\dagger\right>=\int_{0}^\infty dt\;e^{-{C}t}{\Gamma}e^{-{C}^T t}\;,\label{eq:steadyfluc2}
\end{equation}
where
\begin{align}
&\Gamma= \Gamma_{\rm thermal}+\frac{\Gamma_a n_{0,\Sigma}}{R_0^2}\left(\frac{(\ell+2)(\ell-1)R_0\ell(\ell+1)}{2\eta R_0^3(4\ell^3+6\ell^2-1)}\right)^2\times\nonumber\\
&\left(\left(\begin{array}{cc}1&0\\ 0&0\end{array}\right)\left({C}^T+\frac{1}{\tau_\ell}I\right)^{-1}+\left({C}+\frac{1}{\tau_\ell}I\right)^{-1}\left(\begin{array}{cc}1&0\\ 0&0\end{array}\right)\right).
\end{align}
If we assume that $C$ has two distinct eigenvalues $\lambda_1$ and $\lambda_2$ then we can use the Caley-Hamilton theorem to find that we can write the exponential of $-Ct$ as
\begin{align}
e^{-{C}t}=&\frac{1}{\lambda_1-\lambda_2}\big\{\left(\lambda_1 e^{-\lambda_2 t}-\lambda_2 e^{-\lambda_1 t}\right)I\nonumber\\
&+\left(e^{-\lambda_1 t}-e^{-\lambda_2 t}\right){C}\big\}\;.
\end{align}
Using this formula together with
\begin{equation}
\lambda_1+\lambda_2={\rm Tr}\,{C}\;,\quad\lambda_1\lambda_2=\det{C}\;,
\end{equation}
we can perform the integral in Equation (\ref{eq:steadyfluc2}) to find
\begin{align}
\left<\vec{w}_{\ell m}\vec{w}_{\ell m}^\dagger\right>=&\frac{1}{2{\rm Tr}\,{C}\det{C}}\left({\rm Tr}\,{C}\;I-{C}\right){\Gamma}\left({\rm Tr}\,{C}\;I-{C}\right)^T\nonumber\\
&+\frac{1}{2{\rm Tr}\,{C}}{\Gamma}\;.\label{eq:flucform}
\end{align}
Equation (\ref{eq:flucform}) is not singular at $\lambda_1=\lambda_2$ and thus work even if the eigenvalues are identical.

Using Equation (\ref{eq:flucform}) we finally find the spectrum of the shape fluctuations including the contributions from the active force-dipoles
\begin{multline}
\left<|u_{\lmode m}|^2\right>=\frac{k_{\rm B}T}{{\tilde{\tilde{E}}}^e_\ell}\Bigg\{1+\left[\frac{ F_{(2)}\left(F_{(2)}-\Lambda\right) }{\chi} + \frac{\Gamma_a n_{0,\Sigma}}{k_{\rm B}T}x_\ell\right] \\
\cdot \frac{(\ell+2)^2(\ell-1)^2}{4 {\tilde E}_\ell(1+\tau_{\kappa,\ell}/\tau_{D,\ell}) } \Bigg\}\;.\label{eq:anaform}
\end{multline}
Here we have introduced the activity modified energy scales
\begin{align}
{\tilde{E}}_\ell=&\left({\kappa} \ell (\ell+1)+{\tilde\sigma} R_0^2\right)(\ell+2)(\ell-1)\\
{\tilde{\tilde{E}}}^e_\ell=&\left({\tilde{\tilde{\kappa}}}^e \ell (\ell+1)+{\tilde{\tilde{\sigma}}}^e R_0^2\right)(\ell+2)(\ell-1)\;,
\end{align}
where the modified tension and bending parameters are
\begin{align}
{\tilde\sigma}&=\sigma_{0,\phi}+\sigma_{0,{\rm dip}}\;,\\
{\tilde{\tilde{\kappa}}}^e&={\kappa}-\frac{\Lambda(\Lambda-F_{(2)})}{4 \chi}\;,\\
{\tilde{\tilde{\sigma}}}^e&={\tilde\sigma}+\frac{\Lambda(\Lambda-F_{(2)})}{2R_0^2 \chi}\;.
\end{align}
We have also introduced timescales for shape ($\tau_{\kappa,\ell}$) and protein density ($\tau_{D,\ell}$) relaxations 
\begin{align}
\tau_{\kappa,\ell}&=\frac{\eta R_0^3 (4 \ell^3+6 \ell^2-1)}{{\tilde E}_\ell \ell(\ell+1)}\;,\\
\tau_{D,\ell}&=\frac{R_0^2}{D_{\Delta\Delta}\ell(\ell+1)}\;,
\end{align}
where
\begin{equation}
D_{\Delta\Delta}=\Omega_{\Delta\Delta}\chi\label{eq:Deltadiff}
\end{equation}
is the diffusion constant for $n_\Delta$. Finally we have introduced the fraction
\begin{equation}
x_\ell=\frac{{\tilde{\tilde{E}}}^e_\ell+ \frac{\tau_\ell}{\tau_{D,\ell}}\left(1+\frac{\tau_{\kappa,\ell}}{\tau_{D,\ell}}\right){\tilde{E}}_\ell+\frac{\tau_{\kappa,\ell}}{\tau_{D,\ell}}{\tilde{E}}_\ell}{\left(1+\frac{\tau_{\kappa,\ell}}{\tau_{D,\ell}}\right){\tilde{E}}_\ell+\frac{\tau_{\kappa,\ell}}{\tau_{\ell}}{\tilde{E}}_\ell+\frac{\tau_{\ell}}{\tau_{D,\ell}}{\tilde{\tilde{E}}}^e_\ell }\;,\label{eq:xl}
\end{equation}
which determines the importance of the internal active noise.

\begin{figure}
\centerline{
\resizebox{8cm}{!}
{
  \includegraphics{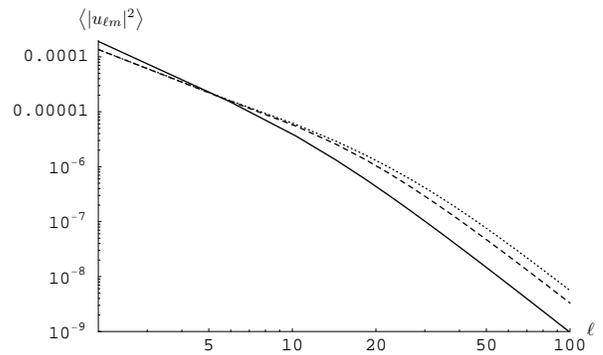}
}}
\caption{The fluctuation spectrum, Equation (\ref{eq:anaform}). The parameters are as explained in the text for the dotted curve. The solid curve is the equilibrium result, i.e. $F_{(2)}$, $\Gamma_a$ and $\sigma_{0,{\rm dip}}$ has been set to zero. For the dashed curve we have taken $\Gamma_a=0$ such that only the internal active noise of the proteins is ignored.}\label{fig:plot1}
\end{figure}

A plot showing how the force-dipole activity modifies the fluctuation spectrum is shown in figure \ref{fig:plot1}. The parameters have been chosen in accordance with the estimates and findings of \cite{manneville01} for a lipid membrane with the proton pump bacteriorhodopsin incorporated. Thus we have taken $\kappa=10 k_{\rm B}T$ with $k_{\rm B}T=4\cdot 10^{-21}\ {\rm J}$. The protein potential $V$ is taken to be purely entropic and thus $\chi=k_{\rm B}T/n_{0,\Sigma}$ with $n_{0,\Sigma}=10^{16}\ {\rm m}^{-2}$ for the total density of proteins. We have assumed that the single protein and collective diffusion constants are identical and taken them to be $D_{\Delta\Delta}=D_{\rm sp}=10^{-12}{\rm m}^2/{\rm s}$. For simplicity we have ignored any asymmetry in the non-active interactions between the proteins and the lipids and taken $\Lambda=0$. We have taken the vesicle to have a radius of $R_0=10\ \mu{\rm m}$ with a tension $\sigma_{0,\phi}=5\cdot 10^{-8}\ {\rm N/m}$ which is increased by 50\% when the proteins are active to ${\tilde \sigma}=7.5\cdot 10^{-8}\ {\rm N/m}$. The relation between the constant $F_{(2)}$ here and the parameters of \cite{manneville01} is that $F_{(2)}=2w\mathcal{P}_a$, where $w$ is the membrane thickness and $\mathcal{P}_a$ has the units of a force-dipole. Using the values of \cite{manneville01} this gives $F_{(2)}=10\,{\rm nm}\cdot\kappa$. The amplitude of the internal noise $\Gamma_a$ should be related to $F_{(2)}$ and here we have simply taken $\Gamma_a=F_{(2)}^2$. For the characteristic time of the active process we have taken the duration of a bacteriorhodopsin photocycle, $\tau_a=5\,{\rm ms}$. Finally the surrounding medium has the viscosity of water $\eta=10^{-3}\ {\rm Pa}\cdot{\rm s}$.

\begin{figure}
\centerline{
\resizebox{8cm}{!}{
  \includegraphics{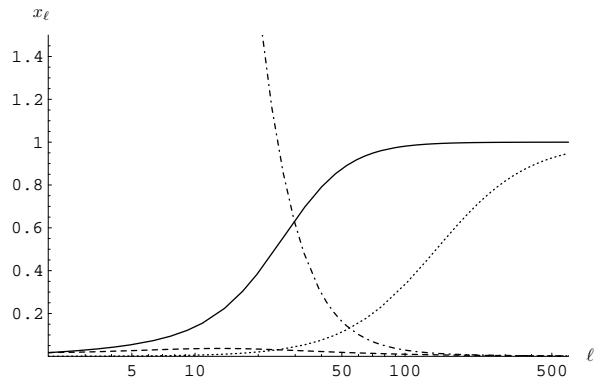}
}}
\caption{The fractions $x_\ell$ (solid curve), $\tau_{\kappa,\ell}/\tau_\ell$ (dash-dotted curve), $\tau_{\kappa,\ell}/\tau_{D,\ell}$ (dashed curve) and $\tau_\ell/\tau_{D,\ell}$ (dotted curve) as functions of $\ell$.}\label{fig:plot2}
\end{figure}

Different ways in which the activity of the force-dipoles modify the fluctuation spectrum can be seen in figure \ref{fig:plot1}. First of all the increase in the tension, $\sigma_{\rm dip}$, decreases the fluctuations at large wavelengths (small $\ell$). Secondly, the thermal fluctuations of $n_\Delta$ enhances the shape fluctuations at short wavelengths through the active contribution to the bending moment: $F_{(2)}n_{\Delta}$. This was the explanation given in \cite{manneville01} for the increase in the fluctuations observed in \cite{manneville99}. However, it can also be seen in the figure that the internal active noise can give a third modification at short wavelength, which sets in gradually as $x_\ell$ increases to become of order one, see figure \ref{fig:plot2}. Note that eventually as the modenumber $\ell$ goes to infinity, the contributions of the activity to the fluctuations scales as $\ell^{-4}$, i.e. in the same way as the thermal fluctuations. In the classification scheme of \cite{gov04,lin06} the wavelength behavior of the fluctuations in the force-dipole model corresponds to the case of ``curvature-force''.

The physics behind the gradual onset of the effect of the internal active noise can be understood from the time-scales of the problem. Using the explicit values of the parameters chosen for the plot we find that at high $\ell$ we have $\tau_{\kappa,\ell}\simeq 100{\rm s}/\ell^3$ and $\tau_{D,\ell}\simeq 100{\rm s}/\ell^2$. Looking at the fractions of time scales in Equation (\ref{eq:xl}) we see that all of these fractions will be at maximum of the order unity except $\tau_{\kappa,\ell}/\tau_\ell$ (see also figure \ref{fig:plot2}), which will be bigger than one for $\ell$ small enough such that $\tau_{\kappa,\ell}$ is larger than $\tau_a$. Since this fraction enters in the denominator of $x_\ell$, we will have that $x_\ell$ is small for $\ell$'s that are smaller than, say, $20$. The physical origin of this is that when the membrane shape dynamics is slower than the dynamics of the internal active noise, then the shape cannot move fast enough to accommodate the temporary stresses that the internal active noise induces. Thus the noise will not influence the movement of the shape at long wavelength.

Note that the timescale for diffusion $\tau_{D,\ell}$ is slower than the timescale for membrane shape relaxation $\tau_{\kappa,\ell}$. A simplifying approximation in Equations (\ref{eq:flucform}) and (\ref{eq:xl}) would thus be to discard terms proportional to  $\tau_{\kappa,\ell}/\tau_{D,\ell}$.

\section{Micropipette experiments}\label{sec:microexp}
To understand what impact including the non-thermal noise represented by $\Gamma_a$ in the force-dipole model has on interpreting existing experimental data, we will briefly discuss micropipette experiments in this section \footnote{The result of the analysis performed in this section consist essentially of summing the appropriate special case of the effective temperature found in \cite{manneville01} with the active temperature found for the case of ``curvature-force'' in \cite{gov04}. It is included here for completeness.}.

In micropipette experiments a relation between the area of a vesicle and the tension in its membrane is measured. The vesicle is aspirated to a pipette at low suction pressure inducing a tension in the membrane of the size $\mu{\rm N}/{\rm m}$ and above \cite{evans90}. The tension is then increased by increasing the suction pressure and it is measured how the visible area of the membrane changes in response. At low tension it is usually assumed that the increase in visible area is taken from the excess area stored in the fluctuations of the membrane without the true area being increased.

We can calculate the excess area, which we will denote by $\alpha'$ to distinguish it from the $\alpha$ of Equation (\ref{eq:Apconstr}), for the force-dipole model applied to an spherical vesicle from Equation (\ref{eq:anaform}). Expanding the area of the vesicle surface, $A$, to second order in $u_{\ell m}$ one finds for a spherical vesicle with fixed volume
\begin{align}
\alpha'&\equiv\frac{A}{4\pi R_0^2}-1\nonumber\\
&=\frac{1}{8\pi}\sum_{\ell\ne 0,m}(\ell+2)(\ell-1)|u_{\ell m}|^2\nonumber\\
&=\frac{1}{8\pi}\sum_{\ell=1}^{\ell_{\rm max}}(2\ell+1)(\ell+2)(\ell-1)|u_{\ell m}|^2\;,\label{eq:excessarea}
\end{align}
where $l_{\rm max}$ is a cut-off wavenumber beyond which the free energy of Equation (\ref{eq:free}) is not expected to be valid anymore. Due to the relatively high tensions in the micropipette experiments (in comparison with what can be achieved in video microscopy experiments) the fluctuations of the low wavenumbers $\ell$ do not contribute much to the excess area. We can therefore simply replace factors of, say, $\ell+2$ in Equations (\ref{eq:anaform}) and (\ref{eq:excessarea}) with $l$ and convert the sum to an integral to simplify the calculation. As mentioned at the end of the last section we can also put factors of $\tau_{\kappa,\ell}/\tau_{D,\ell}$ to zero. To simplify the expression for $x_\ell$ we will simply take $\Lambda$ to be zero such that $x_\ell=1$ for high $\ell$, and assume that the tension is high enough such that the terms in the square brackets of Equation (\ref{eq:anaform}) is suppressed by the tension for the lower $l$ where $x_\ell$ is not unity. With these simplifications we get
\begin{align}
\alpha'&=\frac{k_{\rm B}T^{\rm eff}}{8 \pi \kappa}\log \frac{\kappa l_{\rm max}^2+{\tilde \sigma} R_0^2}{{\tilde \sigma} R_0^2}\nonumber\\
&\simeq -\frac{k_{\rm B}T^{\rm eff}}{8 \pi \kappa}\log \frac{{\tilde \sigma}}{\kappa l_{\rm max}^2/R_0^2}\;,\label{eq:alphasigma}
\end{align}
where
\begin{equation}
\frac{T^{\rm eff}}{T}=1+\frac{F_{(2)}^2}{4\chi\kappa}+\frac{\Gamma_a n_{0,\Sigma}}{4\kappa k_{\rm B}T}\label{eq:Teffform}
\end{equation}
is an effective temperature whose ratio with $T$ is a measure of how much the excess area has been increased by the activity induced fluctuations in comparison with equilibrium.

In \cite{manneville01,girard05} a factor two to three increase of $T^{\rm eff}$ in micropipette experiments on active membranes were reported. This increase was assigned to the averaged force-dipole activity represented by $F_{(2)}$ in Equation (\ref{eq:Teffform}). But as Equation (\ref{eq:Teffform}) shows, the noise of the activity represented by $\Gamma_a$ might well give a contribution of comparable size. Actually, with the estimated values chosen in the previous section the contributions of the corresponding two terms in Equation (\ref{eq:Teffform}) become exactly equal. Further experiments are needed to make it possible to distinguish the contributions of the two mechanisms. An obvious candidate for such experiments is a video microscopy experiment, where the two contributions might be distinguishable due to the onset at high $\ell$-modes of the internal noise represented by $\Gamma_a$ discussed in the last section. However, a limit of the video microscopy experiments is the resolution of the experimental setup, setting a limit for the range of $\ell$-modes that can be obtained in such an experiment. Currently this limit seems to be the first 20 to 30 modes \cite{pecreaux04}, making the possibility of observing an onset very sensitive to the actual parameters, $\kappa$, $\Gamma_a$ etc., of the system.

\section{Conclusion}\label{sec:concl}
In this paper the Gibbs formalism of \cite{lomholt05,lomholt05c} was used to calculate the quasi-spherical shape fluctuation spectrum, given in Equation (\ref{eq:anaform}), for the force-dipole model of \cite{manneville01}. This calculation allows for further testing of the force-dipole through video microscopy experiments, something which is needed to test the model more stringently. The shortcoming of the micropipette experiments is that they only provide one data point for each set of experimental conditions. A video microscopy experiment on the other would provide a data point for each modenumber $\ell$ in the spectrum that can be measured.

Following a general viewpoint promoted in \cite{gov04} an additional contribution due to temporal noise in the strength of the force-dipoles of the individual active proteins was included in the calculation. The contribution of this noise was found to be insignificant at long wavelengths, but it is turned gradually on as the mode number $\ell$ increases when the corresponding characteristic time scale for membrane shape relaxation crosses the time scale associated with the correlations in the noise from the activity. This gradual onset offers a possible way to distinguish the contribution from those of the original force-dipole model of \cite{manneville01} when the shape fluctuation spectrum obtained by for instance a video microscopy experiment is analyzed.

\begin{acknowledgments}
I would like to thank Ling Miao, Per Lyngs Hansen, Tobias Ambj{\"o}rnsson and Jean-Fran{\c c}ois Joanny for many illuminating discussions on active membranes. I am grateful to the Villum Kann Rasmussen Foundation for their financial support.
\end{acknowledgments}

\appendix

\section{Equilibrium fluctuations}\label{sec:appfluc}
For a column vector $\vec{\bar w}_{\ell m}$ with components that perform small fluctuations about an average value zero we can calculate the equilibrium fluctuations as
\begin{equation}
\left<\vec{\bar w}_{\lmode m}\vec{\bar w}_{\lmode m}^\dagger\right>=k_{\rm B}T{\bar A}^{-1}\;,
\end{equation}
where the matrix ${\bar A}$ is the second derivative of the free energy
\begin{equation}
{\bar A}\equiv \left.\frac{\partial^2 F}{\partial \vec{\bar w}_{\ell m}\partial \vec{\bar w}_{\ell m}^\dagger}\right|_{{\bar w}_{\ell m}=0}\;.
\end{equation}

However, for the membrane case studied here there is a subtlety with respect to which fields to use in $\vec{\bar w}_{\ell m}$. When we take the derivative of the free energy $F$ in Equation (\ref{eq:frs}) we fix the fields $\sqrt{g}n_A$ and not $n_A$. This is the reason that we choose for $\vec{\bar w}_{\ell m}$ (see also \cite{miao02})
\begin{equation}
\sum_{\lmode,m}\vec{\bar w}_{\lmode m}\mathcal{Y}_{\ell m}=\left(\begin{array}{c}\frac{1}{R_0}\vec{n}_0\cdot\left(\vec{R}-\vec{R}_0\right)\\ \frac{1}{\sqrt{g_0}}\sqrt{g}n_\Delta\\ \frac{1}{\sqrt{g_0}}\sqrt{g}n_\Sigma-n_{0,\Sigma}\end{array}\right)\;.
\end{equation}
The transformation to the $\vec{w}_{\ell m}$ we have used in the main part of the paper is then at linear order
\begin{equation}
\vec{w}_{\lmode m}=L \vec{\bar w}_{\lmode m}\;,\label{eq:transf}
\end{equation}
where
\begin{equation}
L=\left(\begin{array}{ccc}1&0&0\\0&1&0\\-2n_{0,\Sigma}&0&1\end{array}\right)\;.
\end{equation}

The free energy, Equation (\ref{eq:free}), without the Lagrange multiplier term for $\phi$ but with a tension parameter $\sigma_0'$ included instead to control the area is
\begin{equation}
F=\int_{\rm M} d A\;\left[2\kappa H^2+\Lambda n_\Delta H+V(n_\Delta,n_\Sigma)+\sigma_0'\right]\;.\label{eq:apfree}
\end{equation}

When we expand the free energy to second order in $\vec{\bar w}_{\ell m}$ to find ${\bar A}$ we have to take care of the constraints on volume and number of molecules in the membrane. The volume constraint tells us that
\begin{equation}
V=\frac{1}{3}\int_{\rm M} d A\;\vec{R}\cdot\vec{n}
\end{equation}
should be constant. To second order in $u_{\ell m}$ we have
\begin{equation}
V=R_0^3\left[\frac{4\pi}{3}+\sqrt{4\pi}u_{00}+\sum_{\ell,m}|u_{lm}|^2\right]\;.
\end{equation}
This equation will be used to eliminate $u_{00}$ and express the free energy $F$ in terms of the remaining $u_{\ell m}$ when $F$ is expanded to second order in $u_{\ell m}$. Note that $u_{00}$ depends on $u_{\ell m}$, $\ell\ne 0$ through terms which are second order in $u_{\ell m}$. We will therefore regard $u_{00}$ as a second order term when we expand the free energy.

The constraints on the number of particles, i.e. $\int_{\rm M} d A\;n_A$ is the total number of molecules of type $A$, simply constrains the last two components of $\vec{\bar w}_{00}$ to be zero.

Using the constraints described above we can expand the free energy to second order in $\vec{\bar w}_{\ell m}$ to find
\begin{equation}
F={\rm constant}+\frac{1}{2}\sum_{\lmode\ne 0,m}\vec{\bar w}_{\lmode m}^\dagger{\bar A}_{\lmode m}\vec{\bar w}_{\lmode m}\;,
\end{equation}
where
\begin{multline}
{\bar A}=\\
\left(\begin{array}{ccc}E_\ell+4R_0^2\chi^\Sigma n_{0,\Sigma}^2&-\frac{\Lambda R_0(\ell+2)(\ell-1)}{2}&-2R_0^2\chi^\Sigma n_{0,\Sigma}\\
-\frac{\Lambda R_0(\ell+2)(\ell-1)}{2}&\chi R_0^2&0\\
-2R_0^2\chi^\Sigma n_{0,\Sigma}&0&\chi^\Sigma R_0^2\end{array}\right),
\end{multline}
and the tension that replaces $\sigma_{0,{\rm dip}}$ in the expression for $E_\ell$, Equation (\ref{eq:El}), is
\begin{equation}
\sigma_0=\sigma_0'+\left.V-\frac{\partial V}{\partial n_\Sigma}n_\Sigma-\frac{\partial V}{\partial n_\Delta}n_\Delta\right|_{n_{0,\Delta}=0,n_{0,\Sigma}}\;.
\end{equation}
$\chi^\Sigma$ is simply
\begin{equation}
\chi^\Sigma=\left.\frac{\partial^2 V}{\partial n_\Sigma^2}\right|_{n_{0,\Delta}=0,n_{0,\Sigma}}\;.
\end{equation}

Using the linear transformation, Equation (\ref{eq:transf}), we find that the fluctuations of $\vec{w}_{\ell m}$ with $\ell\ne 0$ are
\begin{equation}
\left<\vec{w}_{\ell m}\vec{w}_{\ell m}^\dagger\right>=k_{\rm B}TA^{-1}\;,
\end{equation}
where
\begin{multline}
A=\left(L^{-1}\right)^T{\bar A}L^{-1}\\
= \left(\begin{array}{ccc}E_\ell&-\frac{\Lambda R_0(\ell+2)(\ell-1)}{2}&0\\
-\frac{\Lambda R_0(\ell+2)(\ell-1)}{2}&\chi R_0^2&0\\
0&0&\chi^\Sigma R_0^2\end{array}\right)\;.
\end{multline}
This is the result used in Equation (\ref{eq:eqfluc}).

\bibliography{../mycites/mycites}

\begin{thebibliography}{17}
\expandafter\ifx\csname natexlab\endcsname\relax\def\natexlab#1{#1}\fi
\expandafter\ifx\csname bibnamefont\endcsname\relax
  \def\bibnamefont#1{#1}\fi
\expandafter\ifx\csname bibfnamefont\endcsname\relax
  \def\bibfnamefont#1{#1}\fi
\expandafter\ifx\csname citenamefont\endcsname\relax
  \def\citenamefont#1{#1}\fi
\expandafter\ifx\csname url\endcsname\relax
  \def\url#1{\texttt{#1}}\fi
\expandafter\ifx\csname urlprefix\endcsname\relax\def\urlprefix{URL }\fi
\providecommand{\bibinfo}[2]{#2}
\providecommand{\eprint}[2][]{\url{#2}}

\bibitem[{\citenamefont{Alberts et~al.}(2002)\citenamefont{Alberts, Johnson,
  Lewis, Raff, Roberts, and Walter}}]{alberts02}
\bibinfo{author}{\bibfnamefont{B.}~\bibnamefont{Alberts}},
  \bibinfo{author}{\bibfnamefont{A.}~\bibnamefont{Johnson}},
  \bibinfo{author}{\bibfnamefont{J.}~\bibnamefont{Lewis}},
  \bibinfo{author}{\bibfnamefont{M.}~\bibnamefont{Raff}},
  \bibinfo{author}{\bibfnamefont{K.}~\bibnamefont{Roberts}}, \bibnamefont{and}
  \bibinfo{author}{\bibfnamefont{P.}~\bibnamefont{Walter}},
  \emph{\bibinfo{title}{Molecular Biology of the Cell}}
  (\bibinfo{publisher}{Garland}, \bibinfo{address}{New York},
  \bibinfo{year}{2002}), \bibinfo{edition}{4th} ed.

\bibitem[{\citenamefont{Seifert}(1997)}]{seifert97}
\bibinfo{author}{\bibfnamefont{U.}~\bibnamefont{Seifert}},
  \bibinfo{journal}{Adv. Phys.} \textbf{\bibinfo{volume}{46}},
  \bibinfo{pages}{13} (\bibinfo{year}{1997}).

\bibitem[{\citenamefont{Manneville et~al.}(1999)\citenamefont{Manneville,
  Bassereau, L\'evy, and Prost}}]{manneville99}
\bibinfo{author}{\bibfnamefont{J.-B.} \bibnamefont{Manneville}},
  \bibinfo{author}{\bibfnamefont{P.}~\bibnamefont{Bassereau}},
  \bibinfo{author}{\bibfnamefont{D.}~\bibnamefont{L\'evy}}, \bibnamefont{and}
  \bibinfo{author}{\bibfnamefont{J.}~\bibnamefont{Prost}},
  \bibinfo{journal}{Phys. Rev. Lett.} \textbf{\bibinfo{volume}{82}},
  \bibinfo{pages}{4356} (\bibinfo{year}{1999}).

\bibitem[{\citenamefont{Girard et~al.}(2005)\citenamefont{Girard, Prost, and
  Bassereau}}]{girard05}
\bibinfo{author}{\bibfnamefont{P.}~\bibnamefont{Girard}},
  \bibinfo{author}{\bibfnamefont{J.}~\bibnamefont{Prost}}, \bibnamefont{and}
  \bibinfo{author}{\bibfnamefont{P.}~\bibnamefont{Bassereau}},
  \bibinfo{journal}{Phys. Rev. Lett.} \textbf{\bibinfo{volume}{94}},
  \bibinfo{pages}{088102} (\bibinfo{year}{2005}).

\bibitem[{\citenamefont{P\'ecr\'eaux et~al.}(2004)\citenamefont{P\'ecr\'eaux,
  D{\"o}bereiner, Prost, Joanny, and Bassereau}}]{pecreaux04}
\bibinfo{author}{\bibfnamefont{J.}~\bibnamefont{P\'ecr\'eaux}},
  \bibinfo{author}{\bibfnamefont{H.-G.} \bibnamefont{D{\"o}bereiner}},
  \bibinfo{author}{\bibfnamefont{J.}~\bibnamefont{Prost}},
  \bibinfo{author}{\bibfnamefont{J.-F.} \bibnamefont{Joanny}},
  \bibnamefont{and}
  \bibinfo{author}{\bibfnamefont{P.}~\bibnamefont{Bassereau}},
  \bibinfo{journal}{Eur. Phys. J. E} \textbf{\bibinfo{volume}{13}},
  \bibinfo{pages}{277} (\bibinfo{year}{2004}).

\bibitem[{\citenamefont{Manneville et~al.}(2001)\citenamefont{Manneville,
  Bassereau, Ramaswamy, and Prost}}]{manneville01}
\bibinfo{author}{\bibfnamefont{J.-B.} \bibnamefont{Manneville}},
  \bibinfo{author}{\bibfnamefont{P.}~\bibnamefont{Bassereau}},
  \bibinfo{author}{\bibfnamefont{S.}~\bibnamefont{Ramaswamy}},
  \bibnamefont{and} \bibinfo{author}{\bibfnamefont{J.}~\bibnamefont{Prost}},
  \bibinfo{journal}{Phys. Rev. E} \textbf{\bibinfo{volume}{64}},
  \bibinfo{pages}{021908} (\bibinfo{year}{2001}).

\bibitem[{\citenamefont{Lomholt}(2006)}]{lomholt05c}
\bibinfo{author}{\bibfnamefont{M.~A.} \bibnamefont{Lomholt}},
  \bibinfo{journal}{Phys. Rev. E} \textbf{\bibinfo{volume}{73}},
  \bibinfo{pages}{061913} (\bibinfo{year}{2006}).

\bibitem[{\citenamefont{Gov}(2004)}]{gov04}
\bibinfo{author}{\bibfnamefont{N.}~\bibnamefont{Gov}}, \bibinfo{journal}{Phys.
  Rev. Lett.} \textbf{\bibinfo{volume}{93}}, \bibinfo{pages}{268104}
  (\bibinfo{year}{2004}).

\bibitem[{\citenamefont{Lomholt et~al.}(2005)\citenamefont{Lomholt, Hansen, and
  Miao}}]{lomholt05}
\bibinfo{author}{\bibfnamefont{M.~A.} \bibnamefont{Lomholt}},
  \bibinfo{author}{\bibfnamefont{P.~L.} \bibnamefont{Hansen}},
  \bibnamefont{and} \bibinfo{author}{\bibfnamefont{L.}~\bibnamefont{Miao}},
  \bibinfo{journal}{Eur. Phys. J. E} \textbf{\bibinfo{volume}{16}},
  \bibinfo{pages}{439} (\bibinfo{year}{2005}).

\bibitem[{\citenamefont{Miao et~al.}(2002)\citenamefont{Miao, Lomholt, and
  Kleis}}]{miao02}
\bibinfo{author}{\bibfnamefont{L.}~\bibnamefont{Miao}},
  \bibinfo{author}{\bibfnamefont{M.~A.} \bibnamefont{Lomholt}},
  \bibnamefont{and} \bibinfo{author}{\bibfnamefont{J.}~\bibnamefont{Kleis}},
  \bibinfo{journal}{Eur. Phys. J. E} \textbf{\bibinfo{volume}{9}},
  \bibinfo{pages}{143} (\bibinfo{year}{2002}).

\bibitem[{\citenamefont{Henriksen and Ipsen}(2004)}]{henriksen04}
\bibinfo{author}{\bibfnamefont{J.~R.} \bibnamefont{Henriksen}}
  \bibnamefont{and} \bibinfo{author}{\bibfnamefont{J.~H.} \bibnamefont{Ipsen}},
  \bibinfo{journal}{Eur. Phys. J. E} \textbf{\bibinfo{volume}{14}},
  \bibinfo{pages}{149} (\bibinfo{year}{2004}).

\bibitem[{\citenamefont{Kim and Karrila}(1991)}]{kim91}
\bibinfo{author}{\bibfnamefont{S.}~\bibnamefont{Kim}} \bibnamefont{and}
  \bibinfo{author}{\bibfnamefont{S.~J.} \bibnamefont{Karrila}},
  \emph{\bibinfo{title}{Microhydrodynamics: Principles and Selected
  Applications}} (\bibinfo{publisher}{Butterworth-Heinemann},
  \bibinfo{address}{Boston}, \bibinfo{year}{1991}).

\bibitem[{\citenamefont{${\rm van}\;{\rm Kampen}$}(1997)}]{kampen97}
\bibinfo{author}{\bibfnamefont{N.~G.} \bibnamefont{${\rm van}\;{\rm Kampen}$}},
  \emph{\bibinfo{title}{Stochastic Processes in Physics and Chemistry}}
  (\bibinfo{publisher}{Elsevier}, \bibinfo{address}{Amsterdam},
  \bibinfo{year}{1997}), \bibinfo{edition}{2nd} ed.

\bibitem[{\citenamefont{Prost and Bruinsma}(1996)}]{prost96}
\bibinfo{author}{\bibfnamefont{J.}~\bibnamefont{Prost}} \bibnamefont{and}
  \bibinfo{author}{\bibfnamefont{R.}~\bibnamefont{Bruinsma}},
  \bibinfo{journal}{Europhys. Lett.} \textbf{\bibinfo{volume}{33}},
  \bibinfo{pages}{321} (\bibinfo{year}{1996}).

\bibitem[{\citenamefont{Prost et~al.}(1998)\citenamefont{Prost, Manneville, and
  Bruinsma}}]{prost98}
\bibinfo{author}{\bibfnamefont{J.}~\bibnamefont{Prost}},
  \bibinfo{author}{\bibfnamefont{J.-B.} \bibnamefont{Manneville}},
  \bibnamefont{and} \bibinfo{author}{\bibfnamefont{R.}~\bibnamefont{Bruinsma}},
  \bibinfo{journal}{Eur. Phys. J. B} \textbf{\bibinfo{volume}{1}},
  \bibinfo{pages}{465} (\bibinfo{year}{1998}).

\bibitem[{\citenamefont{Lin et~al.}(2006)\citenamefont{Lin, Gov, and
  Brown}}]{lin06}
\bibinfo{author}{\bibfnamefont{L.~L.-C.} \bibnamefont{Lin}},
  \bibinfo{author}{\bibfnamefont{N.}~\bibnamefont{Gov}}, \bibnamefont{and}
  \bibinfo{author}{\bibfnamefont{F.~L.~H.} \bibnamefont{Brown}},
  \bibinfo{journal}{J. Chem. Phys.} \textbf{\bibinfo{volume}{124}},
  \bibinfo{pages}{074903} (\bibinfo{year}{2006}).

\bibitem[{\citenamefont{Evans and Rawicz}(1990)}]{evans90}
\bibinfo{author}{\bibfnamefont{E.}~\bibnamefont{Evans}} \bibnamefont{and}
  \bibinfo{author}{\bibfnamefont{W.}~\bibnamefont{Rawicz}},
  \bibinfo{journal}{Phys. Rev. Lett.} \textbf{\bibinfo{volume}{64}},
  \bibinfo{pages}{2094} (\bibinfo{year}{1990}).

\end{thebibliography}

\end{document}